\documentclass[conference,10pt]{IEEEtran}

\usepackage{setspace}

\usepackage{amsmath}
\usepackage{amssymb}
\usepackage{cite}
\usepackage{graphicx}
\usepackage{epstopdf}
\usepackage{url}
\usepackage{cite}
\usepackage{texdef2015}

\newtheorem{lemma}{Lemma}

\newcommand{\wk}[1][k]{\omega_{#1}}

\newcommand{\nxtic}{ ++(10,0) -- +(0,-1)}
\newcommand{\nytic}{ ++(0,10) -- +(-1,0)}

\newcommand{\Xk}[1][k]{X^{(#1)}}
\newcommand{\Pk}[1][k]{P^{(#1)}}

\newcommand{\mutil}{\tilde{\mu}}
\newcommand{\muhat}{\hat{\mu}}
\newcommand{\age}{\Delta}
\newcommand{\LX}{L}
\newcommand{\lk}[1][k]{\ell_{#1}}
\newcommand{\Fk}[2][k]{F_{#1}(#2)}
\newcommand{\FRage}[1]{\age_{\text{FR}}(#1)}
\newcommand{\FRageOpt}{\age_{\text{FR}}^*}
\newcommand{\ageIIR}{\age_{\text{IIR}}}
\newcommand{\FRUB}[1]{\age_{\overline{\text{FR}}}(#1)}

\newcommand{\negfigspace}{\vspace{-4mm}}

\usepackage{tikz}
\usetikzlibrary{automata,arrows,positioning,calc,fit,shapes.multipart,chains,shapes}
\usepackage{microtype}

\begin{document}

\title{Timely Updates over an Erasure Channel}
\author{$^\dag$Roy D.~Yates, $^\ddag$Elie Najm, $^\dag$Emina Soljanin, and $^{\dag}$Jing Zhong,\\
\small $^\dag$ ECE Dept., Rutgers University, \{ryates,  emina.soljanin, jing.zhong\}@rutgers.edu,
$^{\ddag}$LTHI, EPFL, elie.najm@epfl.ch}

 \maketitle
\begin{abstract} Using an age of information (AoI) metric, we examine the  transmission of  coded updates through a binary erasure channel to a monitor/receiver.
We start by deriving the average status update age of an infinite incremental redundancy (IIR) system in which the transmission of a $k$-symbol update continues until $k$  symbols are received. This system is then compared to a fixed redundancy (FR) system in which each update is transmitted as an $n$ symbol packet and the packet is successfully received if and only if  at least $k$ symbols are received. If fewer than $k$ symbols are received, the update is discarded. Unlike the IIR system, the FR system requires no feedback from the receiver. For a single monitor system, we show that tuning the redundancy to the symbol erasure rate enables the FR system to perform as well as the IIR system.  As the number of monitors is increased, the FR system outperforms the IIR system that guarantees delivery of all updates to all monitors. \end{abstract}
\section{Introduction}
Status update systems have focused on sending updates to a monitor through a system or network in which the transmission of an update requires a random service time \cite{KaulYatesGruteser-2012Infocom,CostaCodreanuEphremides2016,Kam-PathDiversity2016,%
Yates-2015ISIT,UpdateorWait-2016Infocom,UpdateorWait-2016arxiv}. In this work, we consider a system in which a source sends coded updates through an unreliable channel to a monitor.  We examine how to employ coding redundancy in order to minimize an ``Age of Information'' (AoI) metric.  We will see that this is not the same as ensuring reliable delivery of every update while minimizing the coding overhead. Over a noisy channel, the age is reduced only when an update is correctly received, but using coding to increase the probability of correct reception incurs additional delay for each update. In this work, we examine what is just the right amount of redundancy. 

In our system model, 
the source sends updates through a symbol erasure channel to a monitor. One symbol is transmitted per unit time.  A symbol is erased with probability $\delta$;  otherwise it is received correctly. Each update is a timestamped file with $k$ information symbols. Depending on the coding strategy, these updates are encoded into 
 at least $k$ and possibly infinitely many (rateless) coded symbols for transmission over the channel.  The monitor may employ  a feedback channel to notify the source about symbols that have been erased or packets that have failed to be decoded. 

If at time $t$, the most recent received update is timestamped $u(t)$, the status age is $\age(t)=t-u(t)$. In the absence of an update, the status age increases linearly with time. Thus the age process $\age(t)$ is a sawtooth waveform as shown in Figure~\ref{fig:ageIIR}. 
To compare coding strategies, our system performance metric is the time-average status age (also known as the AoI)
\begin{align}\eqnlabel{average-age}
\age&=\limty{T}\frac{1}{T}\int_0^T\age(t)\,dt.
\end{align}

We will use two coding techniques: 1) an infinite incremental redundancy  (IIR) strategy
and 2) a finite redundancy  (FR) strategy.  
Under the IIR strategy,  each $k$-symbol update is encoded by  a rateless code such that when $k$ coded symbols are correctly received by the monitor, the update is successfully decoded (e.g., a Reed-Solomon or a Fountain code). The source is provided instantaneous feedback when the update has been decoded, at which point it starts transmitting a new update. 

Under the FR strategy, each $k$-symbol update is encoded as an $n$-symbol packet. The update is successfully delivered as soon as $k$ un-erased symbols are received. If fewer than $k$ symbols are received, the update is discarded. This system employs no feedback from the monitor and thus all $n$ symbols of an update are transmitted even if the monitor successfully decodes the update before the transmission is finished. The source starts transmitting a new update once the $n$ symbols of the previous update have been sent. 
Note that under the FR strategy 1) not every update will be decoded and 2) there generally will be a positive time gap between the completion of decoding of an update and the beginning of the new update transmission. 


In this work, Section~\ref{sec:IIR} analyzes AoI for the IIR system, first with  a single monitor and then with  $m>1$ monitors. In Section~\ref{sec:FR}, we characterize age in the FR updating system. 
For this system, 
we show that by matching the redundancy $n$ to the erasure rate,  the FR system has AoI approaching that of the single-monitor IIR system as $k$ becomes large. In Section~\ref{sec:eval}, we present numerical evaluations of both systems. A brief discussion concludes this work in Section~\ref{sec:conclusion}.

 \section{AoI Under the IIR Strategy}\label{sec:IIR}
\subsection{Single Monitor System}
Update $1$ begins transmission at time $t=0$ and is timestamped $T_0=0$. To analyze the average age, we define $X_i$ as the number of symbols sent until the $k$th un-erased symbol of update $i$ is received.
Because the erasure channel is memoryless, $X_1,X_2,\ldots$ are iid negative binomial (NB) $(k,1-\delta)$ random variables, identical to $\Xk$  with PMF 
\begin{align}\eqnlabel{X-PMF}
\pmf{\Xk}{x}=\binom{x-1}{k-1}(1-\delta)^k\delta^{x-k},\quad x=k,k+1,\ldots
\end{align}
For convenience, we will denote the CDF of $\Xk$ by
\begin{align}\eqnlabel{X-CDF}
\Fk{n}=\sum_{x=k}^n\binom{x-1}{k-1}(1-\delta)^k\delta^{x-k},\  n=k,k+1,\ldots
\end{align}
We also note that $\Xk$ has expected value $\E{\smash{\Xk}}=\mu_k$ and variance $\Var{\smash{\Xk}} =\sigma_k^2$ with
\begin{align}\eqnlabel{Xk-mean-var}
\mu_k&=\frac{k}{1-\delta}, & \sigma_k^2=\frac{k\delta}{(1-\delta)^2}.
\end{align}
Following the delivery of update $l$ at time $T_{l}=\sum_{i=1}^l X_l$, update $l+1$ immediately begins transmission.

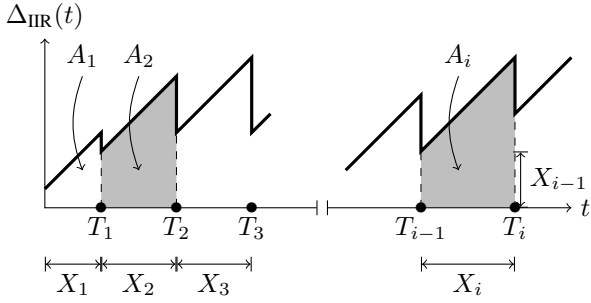
\begin{figure}[t]
\centering
\begin{tikzpicture}[scale=0.25]
\draw [fill=lightgray, ultra thin, dashed] (3,0) to (3,3) to (7,7) to (7,0);
\draw [fill=lightgray, ultra thin, dashed] (20,0) to (20,3) to (25,8) to (25,0);
\draw [<-|] (0,9) node [above] {$\ageIIR(t)$} -- (0,0) -- (14.5,0);
\draw [|->] (15,0) -- (28,0) node [right] {$t$};
\draw 
(3,0) node {$\bullet$} node [below] {$T_1$} 
(7,0) node {$\bullet$} node [below] {$T_2$} 
(11,0) node {$\bullet$} (11,0) node [below] {$T_3$} 
(20,0) node {$\bullet$} node [below] {$T_{i-1}$} 
(25,0) node {$\bullet$} node [below] {$T_{i}$};
\draw [|<->|] (25.3,0) to node [right] {$X_{i-1}$} (25.3,3); 
\draw  [<-] (2,2) to [out=110,in=250] (2,7) node [above] {$A_1$};
\draw[<-] (5,2) to [out=110,in=250] (5,7) node [above] {$A_2$};
\draw[<-] (22,2) to [out=110,in=250] (22,7) node [above] {$A_{i}$};
\draw [very thick] (0,1) -- (3,4) -- (3,3)  -- (7,7) 
-- (7,4)  -- (11,8)-- (11,4) --(12,5);
\draw [very thick] (16,2) -- (20,6) -- (20,3)  -- (25,8) -- (25,5) -- (28,8); 
\draw  [|<->|] (0,-3) to node [below] {$X_1$} (3,-3);
\draw  [|<->|] (3,-3) to node [below] {$X_2$} (7,-3);
\draw  [|<->|] (7,-3) to node [below] {$X_3$} (11,-3);
\draw  [|<->|] (20,-3) to node [below] {$X_{i}$} (25,-3);
\end{tikzpicture}
\caption{Sample path of the status update age $\ageIIR(t)$ (the upper envelope in bold) for the IIR updating system.  Updates are delivered at time instances marked by $\bullet$. Update $i$ submitted at time $T_{i-1}$ is delivered after a transmission time $X_i$.}
\label{fig:ageIIR}\negfigspace
\end{figure}

To analyze the average age $\age$, we decompose the area defined by the integral \eqnref{average-age} into a sum of disjoint polygonal areas $A_1,A_2,\ldots$ as shown in Figure~\ref{fig:ageIIR}. Over the time interval $(0,T=T_l)$, this decomposition yields average age
\begin{align}
\ageIIR = \limty{l}\frac{1}{T_l}\sum_{i=1}^{l} A_i
=\limty{l}\frac{\frac{1}{l}\sum_{i=1}^{l} A_i}{\frac{1}{l}\sum_{i=1}^{l} X_i}=\frac{\E{A}}{\E{X}}.
\eqnlabel{ageIIR1}
\end{align}
When update $i$ begins transmission at time $T_{i-1}$, the age is $\age(T_{i-1})=X_{i-1}$. From Figure~\ref{fig:ageIIR}, we see that the area $A_i$ is
\begin{align}
A_i=X_{i-1}X_i +X_i^2/2.
\end{align}
Since the $X_i$ are iid, $\E{A}=(\E{X})^2+\E{X^2}/2$ and it follows from \eqnref{Xk-mean-var} and \eqnref{ageIIR1} that the average age of the IIR system is
\begin{align}\eqnlabel{ageIIR}
\ageIIR=\E{X}+\frac{\E{X^2}}{2\E{X}}
=\frac{k}{1-\delta}\paren{\frac{3}{2}+\frac{\delta}{k}}.
\end{align}
We note IIR is the only strategy that guarantees the delivery of every update. Moreover, it minimizes the coding overhead, and thus maximizes the throughput. It takes $k/(1-\delta)$ coded symbols on average to transmit a $k$-symbol update, which is not equal to the average update age.
In particular, $3k/[2(1-\delta)]$ is
 what the average age would be if each update were delivered by exactly $k/(1-\delta)$ symbol transmissions. The additional (though admittedly small) age penalty of IIR reflects the randomness in the negative binomial distribution.

We also observe that IIR is a zero-wait system: as soon as an update is delivered, a new update goes into service. However, when service times are random, zero-wait policies may not be age-minimizing. By\cite[Theorem~5]{UpdateorWait-2016arxiv}, it can be shown that  zero-wait  is optimal for IIR if and only if $\delta\le k/(2k+1)$. 

\subsection{Multiple Monitor System}
Using IIR to transmit to $m>1$ monitors, the source continues to transmit encoded symbols until each of the $m$ monitors has correctly received $k$ coded symbols. The source is provided instantaneous feedback when an update has been decoded by all users.

Update $1$ begins transmission at time $t=0$ and is timestamped $T_0=0$. To analyze the average age, we define $X_{ij}$ as the number of symbols sent until the $k$th un-erased symbol of update $i$ is received by monitor $j$.
Because the erasure channels of all users are memoryless and independent, the $X_{ij}$ are iid NB $(k,1-\delta)$ random variables with PMF given by \eqnref{X-PMF}.

The transmission time of update $i$ is
\begin{align}
Y_i&=\max(X_{i1}, \ldots, X_{im}).
\end{align}
The $Y_i$ are an  iid sequence, each with CDF
\begin{align}\eqnlabel{Y-CDF}
\icdf{Y}=\prob{Y\le y}=\prob{\smash{\Xk}\le y}^m=[\Fk{y}]^m.
\end{align}
Following the delivery of update $l$ at time $T_{l}=\sum_{i=1}^l Y_{i}$, update $l+1$ immediately begins transmission. 
\begin{figure}[t]
\centering
\begin{tikzpicture}[scale=0.25]
\draw [fill=lightgray, ultra thin, dashed] (0,0) to (0,4) to (3,7) to (3,3) to (7,7) to (7,0);
\draw [fill=lightgray, ultra thin, dashed] (17,0) to (17,6) to (20,9) to (20,3) to (23,6) to (23,0);
\draw [<-|] (0,12) node [above] {$\ageIIR^{(m)}(t)$} -- (0,0) -- (13,0);
\draw [|->] (14,0) -- (28,0) node [right] {$t$};
\draw 
(3,0) node {$\bullet$}  
(7,0) node [below] {$T_1$}
(11,0) node {$\bullet$}
(17,0) node [below] {$T_{i-1}$} 
(20,0) node {$\bullet$} 
(23,0) node [below] {$T_{i}$}
(25,0) node {$\bullet$};
\draw  [<-] (3,2) to [out=60,in=290] (5,7) node [above] {$A_1$};
\draw[<-] (20,2) to [out=60,in=280] (22,7) node [above] {$A_{i}$};
\draw [very thick] (0,4) -- (3,7) -- (3,3)  -- (7,7) 
-- (11,11)-- (11,4) --(13,6);
\draw [very thick] (14,3) -- (20,9) -- (20,3)  -- (25,8) -- (25,5) -- (28,8); 
\draw  [|<->|] (0,10) to node [above] {$Y_1$} (7,10);
\draw  [|<->|] (17,10) to node [above] {$Y_i$} (23,10);
\draw  [|<->|] (0,-2.5) to node [below] {$X_{1j}$} (3,-2.5);
\draw  [|<->|] (7,-2.5) to node [below] {$X_{2j}$} (11,-2.5);
\draw  [|<->|] (17,-2.5) to node [below] {$X_{ij}$} (20,-2.5);
\end{tikzpicture}
\caption{Sample path of the status update age $\ageIIR^{(m)}(t)$ for user $j$ in the IIR updating system with $m>1$ monitors. Update $i$ completes transmission at time $T_i$. Update delivery instances for monitor $j$ are marked by $\bullet$.}
\label{fig:IIR-muser}\negfigspace
\end{figure}
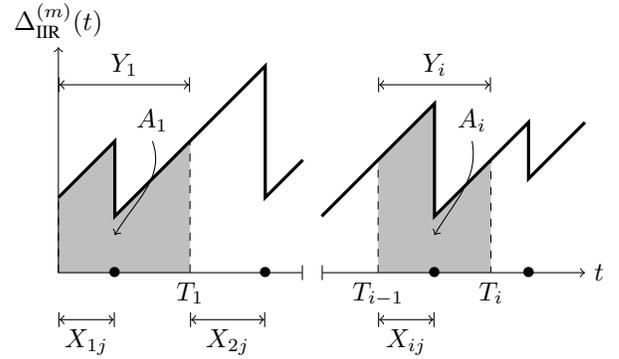  
Since all monitors have statistically identical (but independent) channels, we define $\age(t)$ as the age of some monitor $j$ and we now analyze the average age $\age$. Figure~\ref{fig:IIR-muser} depicts the age process $\age(t)$ for monitor $j$. 
The analysis of the average age is similar to that for the single user IIR system. As before, update $i$ completes transmission at time $T_i$, but,  for user $j$, the age $\age(t)$ drops when update $i$ is delivered to monitor $j$ at the earlier time $T_{i-1}+X_{ij}$. We note this implies that update $i$ completes transmission at time $T_i$, the age at monitor $j$ is then $\age(T_i)=Y_i$. 

As we did for the single user system, we represent the area of the integral \eqnref{average-age} as the concatenation of the polygons $A_1,\ldots,A_l$,  yielding  the average age
$\ageIIR^{(m)}=\E{A}/\E{Y}$.
Examination of Figure~\ref{fig:IIR-muser} will show that
\begin{align}
A_i&=Y_{i-1}X_{ij} +X_{ij}^2/2 +X_{ij}(Y_i-X_{ij})+(Y_i-X_{ij})^2/2\nn
&=Y_{i-1}X_{ij}+Y_i^2/2.
\end{align}
Since  $X_{ij}$ is independent of  the transmission time $Y_{i-1}$ of the previous update, $\E{A}=\E{Y}\E{X}+\E{Y^2}/2$ and
\begin{align}\eqnlabel{ageIIR-m}
\ageIIR^{(m)}=\E{X}+\frac{\E{Y^2}}{2\E{Y}}.
\end{align}
Using \eqnref{X-CDF} and \eqnref{Y-CDF}, the moments $\E{Y}$ and $\E{Y^2}$ are easy to calculate but they do not have simple closed  form expressions. 

\section{AoI under Fixed Redundancy Coding}\label{sec:FR}
Under the fixed redundancy (FR) strategy,  each update is encoded as an $n$-symbol packet but the update is successfully decoded as soon as $k$ un-erased symbols are received. If fewer than $k$ symbols are received, the update is discarded. This system employs no feedback from the monitor and thus all $n$ symbols of an update are transmitted even if the monitor  decodes the update before the transmission is finished.
  
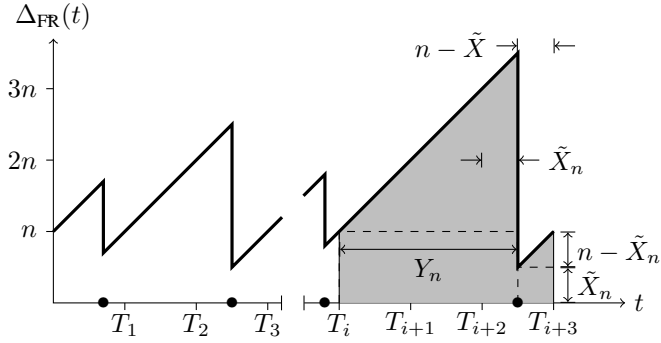
\begin{figure}[t]
\centering
\begin{tikzpicture}[scale=0.095]
\draw [<-|] (0,37) node [above] {$\FRage{t}$} -- (0,0) -- (32,0);
\draw [|->] (35,0) -- (80,0) node [right] {$t$};
\draw (0,0) \nxtic\nxtic\nxtic\nxtic\nxtic \nxtic \nxtic;
\draw (0,0) \nytic\nytic\nytic\nytic;
\draw [fill=lightgray] (40,0) to ++(0,10) to ++(25,25) to ++(0,-30)  to ++(5,5) to ++(0,-10);
\draw (7,0) node {$\bullet$}
(25,0) node {$\bullet$}
(38,0) node {$\bullet$}
(65,0) node {$\bullet$};
\draw (10,-6) node [above] {$T_1$} 
(20,-6) node [above] {$T_2$} (30,-6) node [above] {$T_3$} (40,-6) node [above] {$T_{i}$} (50,-6) node [above] {$T_{i+1}$} (60,-6) node [above] {$T_{i+2}$} (70,-6) node [above] {$T_{i+3}$};
\draw (-1,10) node [left] {$n$} (-1,20) node [left] {$2n$} (-1,30) node [left] {$3n$};
\draw [very thick] (0,10) -- ++(7,7) -- ++(0,-10) -- ++(3,3) -- ++(15,15)  -- ++(0,-20) -- ++(7,7);
\draw [very thick] (35,15) -- ++(3,3) -- ++(0,-10) -- ++(2,2) -- ++(25,25) -- ++(0,-30) -- ++(5,5);
\draw  [|<->|] (40,7.5) to node [below] {$Y_n$} (65,7.5);
\draw [|<-] (65,36) to (62,36) node [left] {$n-\Xtil$};
\draw [|<-] (70,36)  to (73,36) ;
\draw [|<-] (60,20) to (57,20);
\draw [|<-] (65,20) to (68,20) node [right] {$\Xtil_n$};
\draw  [|<->|] (72,0) to node [right] {$\Xtil_n$} (72,5);
\draw  [|<->|] (72,5) to node [right] {$n-\Xtil_n$} (72,10);
\draw [thin,dashed] (40,0) -- (40,10) -- (65,10);
\draw [thin,dashed] (65,0) --(65,5) -- ( 70,5);
\end{tikzpicture}
\caption{\small A sample path of the  FR age $\FRage{t}$: successful update deliveries (at times marked by $\bullet$) occur in slots $1$, $3$, $i$, and $i+3$. Updates are discarded in slots $2$, $i+1$, and $i+2$.}
\label{fig:agesimple}
\negfigspace
\end{figure}

To analyze this system, we define $T_i=in$ and slot $i$ as the time interval $(T_{i-1},T_i]$.
Update $i$ is successfully delivered in  slot $i$ with probability
\begin{align}
1-\epsilon_n =\prob{\Xk\le n} =\Fk{n}.
\eqnlabel{XkCDF}
\end{align}
Referring to Figure~\ref{fig:agesimple}, when a success occurs in slot $i$, the age at time $T_{i-1}+X_i$ is reset to $X_i$  because update $i$ has age  $X_i$ at that time  instant.
Moreover, $\age(T_i)=n$ because update $i$ will then have age $n$ at the end of slot $i$.
Consequently, when a  success occurs in slot $i$, $T_i$ is a renewal point of the process $\age(t)$ in that $\age(T_i)=n$ and time instant $T_i$ marks the start of  transmission of a fresh update.  In the example of Figure~\ref{fig:agesimple}, renewals occur at times $T_1$, $T_3$, $T_i$, and $T_{i+3}$.

Measured in slots, the  length of a renewal period is a geometric $(1-\epsilon_n)$ random variable $M_n$ with PMF
\begin{align}
\pmf{M_n}{m}=\epsilon_n^{m-1}(1-\epsilon_n),\qquad m=1,2,\ldots,
\end{align}
corresponding to $m-1$ updates being discarded followed by a success with update $m$. 

For the FR system, we analyze the AoI $\age$ in \eqnref{average-age} using renewal-reward theory \cite{gallager2013stochastic}. Specifically, we interpret $\age(t)$ as an instantaneous reward rate so that $\age$ is the average reward rate. In the renewal period starting at time $T_i$,  the reward 
\begin{align}
R=\int_{T_i}^{T_{i+M_n}}\age(t)\,dt
\end{align}
is earned. In Figure~\ref{fig:agesimple}, $R$ is the shaded area.  
This renewal period  terminates after $M_n=m$ slots  because $X_{i+m}\le n$. This implies $X_{i+M_n}$ is identical to a random variable $\Xtil_n$ with PMF $\pmf{\Xtil_n}{x}=\pmf{\Xk|\Xk\le n}{x}$. From \eqnref{X-PMF} and \eqnref{X-CDF},
\begin{align}
\pmf{\Xtil_n}{x}
&=\frac{\binom{x-1}{k-1}(1-\delta)^k\delta^{x-k}}{\Fk{n}},\quad
x\le n.
\eqnlabel{Xtildefn}
\end{align}

It will be convenient to define $\mutil_n\equiv\E{\smash{\Xtil_n}}$ and we note that \eqnref{Xtildefn} implies
\begin{align}
\mutil_n &=\frac{1}{\Fk{n}}\sum_{x=k}^n x\binom{x-1}{k-1} (1-\delta)^k\delta^{x-k}\nn
&=\frac{k}{\Fk{n}}\sum_{x=k}^n \binom{x}{k} (1-\delta)^k\delta^{x-k}.
\end{align}
With the substitutions $x'=x+1$ and $k'=k+1$, we obtain
 \begin{align}
\mutil_n
&=\frac{k}{(1-\delta)\Fk{n}}\sum_{x'=k'}^{n+1} \binom{x'-1}{k'-1} (1-\delta)^{k'}\delta^{x'-k'}\nn
&=\frac{k\Fk[k+1]{n+1}}{(1-\delta)\Fk{n}}.\eqnlabel{mutil}
\end{align}

Note that $\Xtil_n$ is independent of the number of slots $M_n$ in a given renewal period.
Referring to Figure~\ref{fig:agesimple}, the renewal period consists of an interval of length
\begin{equation}\eqnlabel{Ydefn}
Y_n=n(M_n-1)+\Xtil_n=nM_n -(n-\Xtil_n).
\end{equation}
in which $\age(t)$ grows from $\age(T_j)=n$ to $\age(T_j+Y)=n+Y_n$, followed by a second interval of length $n-\Xtil_n$. As shown in the figure, each of these intervals contributes a  rectangular area and a triangular area  to the reward $R$.  Thus,
\begin{align}
R&=nY_n+Y_n^2/2 +\Xtil_n(n-\Xtil_n) +(n-\Xtil_n)^2/2\nn
&=nY_n+Y_n^2/2+n^2/2 -\Xtil_n^2/2.
\end{align}
It then follows from \eqnref{Ydefn} that
\begin{align}
R &= n^2M_n^2/2 + nM_n\Xtil_n.
\end{align}
Since the renewal period has length $M_nn$, the renewal-reward theorem ensures that the time-average reward (corresponding to the time-average age $\age$) is
\begin{align}
\FRage{n}=\frac{\E{R}}{\E{M_nn}}
=\frac{n\E{M_n^2}}{2\E{M_n}} +\mutil_n.
\end{align}

Since $M_n$ has moments
\begin{align}
\E{M_n}&=\frac{1}{1-\epsilon_n}, &
\E{M_n^2}&=\frac{1+\epsilon_n}{(1-\epsilon_n)^2},
\end{align}
it follows that 
\begin{align}\eqnlabel{FRage1}
\FRage{n}
&=\frac{n}{1-\epsilon_n} -\frac{n}{2} +\mutil_n.
\end{align}
We note that calculation of $\FRage{n}$ is straightforward using \eqnref{XkCDF} and \eqnref{mutil}.
\subsection{AoI Bounds under FR}
We will see from numerical evaluations in Section~\ref{sec:eval} that  given $k,\delta$ there exists an optimal redundancy $n^*_k$ such that
\begin{align}\eqnlabel{FRbounds}
\FRageOpt=\FRage{n^*_k}\le\FRage{n}
\end{align}
for all $n$.  
To characterize $n^*_k$, we now derive $\FRUB{n}$, a surprisingly tight upper bound on the average age $\FRage{n}$. We then show that a close approximation to $n^*_k$ can be found by a minimization of  $\FRUB{n}$ based on a central limit theorem (CLT) approximation.  While this method is approximate, the result will yield a strict (and tight) upper bound to $\FRageOpt$. 
We start with the following claim, with proof  in the Appendix.
\begin{lemma}\label{lem:rn} For fixed $k$ and $\delta$, the sequence $\mutil_k,\mutil_{k+1},\ldots$ is nondecreasing and satisfies
\[\mutil_n\le \min\paren{n,k/(1-\delta)}.\]
\end{lemma}
%
Applying Lemma~\ref{lem:rn} to \eqnref{FRage1}, we obtain the upper bound
\begin{align}
\FRage{n}\le \FRUB{n}\equiv \frac{n}{1-\epsilon_n}-\frac{n}{2} 
+\frac{k}{1-\delta}.\eqnlabel{AgeQ-UB}
\end{align}



Writing $n=\sigma_kz+\mu_k$, we employ the CLT approximation
\begin{align}
1-\epsilon_n&=\prob{\Xk\le n}\nn
&=\prob{\Xk\le \sigma_kz+\mu_k} \approx \Phi(z)
\end{align}
where $\Phi(z)$ is the standard Gaussian CDF.  Applied to \eqnref{AgeQ-UB}, this approximation permits us to write 
\begin{align}
\FRUB{n}&\approx \frac{\sigma_kz+\mu_k}{\Phi(z)} -\frac{\sigma_kz+\mu_k}{2} +\mu_k\nn
&=\sigma_k\paren{\frac{z+\muhat_k}{\Phi(z)}-\frac{z}{2}+\frac{\muhat_k}{2}}
\end{align}
where 
$\muhat_k\equiv\mu_k/\sigma_k=\sqrt{k/\delta}$.
For large $k$, we will want the probability an update is decoded to be fairly close to $1$. Hence $\Phi(z)\approx 1$ and
$\frac{z}{\Phi(z)}\approx z$ for values of $z$ of interest. 
Thus we make the further approximation
\begin{align}
\FRUB{n}&\approx \sigma_k\paren{\frac{\muhat_k}{\Phi(z)}+\frac{z}{2}+\frac{\muhat_k}{2}}.\eqnlabel{FRUB-v3}
\end{align}
Setting the derivative of the right side of \eqnref{FRUB-v3} to zero, we obtain 
$-\muhat_k\Phi'(z)=[\Phi(z)]^2/2$. 
Using the fact that $\Phi(z)\approx 1$ for values of $z$ of interest and 
since $\Phi'(z)=e^{-z^2/2}/\sqrt{2\pi}$,  solving $\Phi'(z)=-1/[2\muhat_k]$ yields
$z=z^*_k= \sqrt{\ln(2k/\pi\delta)}$.
Employing \eqnref{Xk-mean-var},  we obtain the 
threshold
\begin{align}
\nhat^*_k=\mu_k+\sigma_kz^*_k
&=\frac{k}{1-\delta}\paren{1+\wk }
\end{align}
where
\begin{align}\eqnlabel{ekdefn}
\wk\equiv\bracket{\frac{\delta}{k}\ln\frac{2k}{\pi\delta}}^{\mathrlap{1/2}}.
\end{align}

\begin{figure}[t]
\centering
\includegraphics{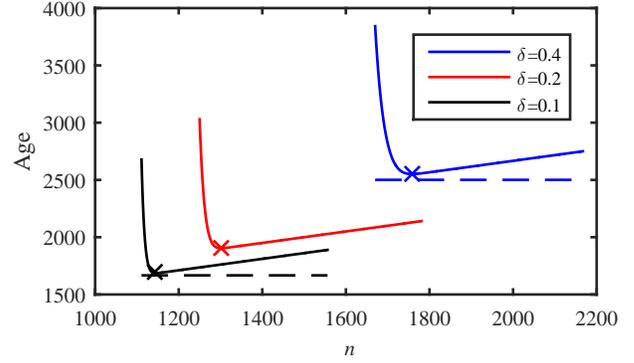}
\caption{The average age $\FRage{n}$ for update packets with $k=1000$ information symbols as a function of the number of transmitted symbols $n$. For each curve, $\times$ marks $\FRUB{\nhat^*_k}$. For $\delta=0.4$ and $\delta=0.1$, the dashed lines show the IIR age $\ageIIR$.}\label{fig:FR1000}\negfigspace
\end{figure}

In the Appendix, we verify the following claim:
\begin{lemma}\label{lem:Xktail} Given $\eta_0>0$, there exists $K_0$ such that
\[\prob{\Xk> \nhat^*_k}\le \beta_k \equiv e^{\eta_0/(1-\delta)}\sqrt{\frac{\pi\delta}{2k}},\qquad k\ge K_0.\]
\end{lemma}
We note that the tail probability $[\prob{\Xk> \nhat^*_k}$ decays slowly (i.e. sub-exponentially) because  $\nhat^*_k$ is approaching $\mu_k$ as $k$ becomes large. 
It follows from \eqnref{FRbounds}, \eqnref{AgeQ-UB} and Lemma~\ref{lem:Xktail} that
\begin{align}
\FRageOpt
\le\FRUB{\nhat^*_k}\
&= \frac{\nhat^*_k}{\prob{\Xk\le\nhat^*_k}}-\frac{\nhat^*_k}{2}+\frac{k}{1-\delta}\nn
&=\frac{k}{1-\delta}\bracket{\frac{3}{2}+\frac{\beta_k +\frac{1}{2}\wk(1+\beta_k)}{1-\beta_k}}.\eqnlabel{FRopt}
\end{align}
Since $\beta_k$ and $\wk$ approach zero as $k$ grows,  we see from \eqnref{ageIIR} and \eqnref{FRopt} that the average age of the IIR system and the average age of FR system with optimized redundancy both asymptotically approach $1.5k/(1-\delta)$.

\section{Evaluation}\label{sec:eval}
Figure~\ref{fig:FR1000} evaluates a system in which updates have $k=1000$ information symbols. We plot the FR  age $\FRage{n}$ in \eqnref{FRage1} as a function of $n$, the FR update packet length, for a range  of values of the erasure probability $\delta$. As one would expect, the age increases as $\delta$ increases. We also see that for a given erasure probability $\delta$, the optimal $n$ is sharply defined. Too few transmitted symbols and the age blows up because the packet update erasure probability is high; on the other hand, more than the minimum number of sent symbols also creates unnecessary age. Marked by $\times$ are the approximately optimal redundancy $\nhat^*_k$  and the corresponding age upper bound $\FRUB{\nhat^*_k}$. 

\begin{figure}[t]
\centering
\includegraphics{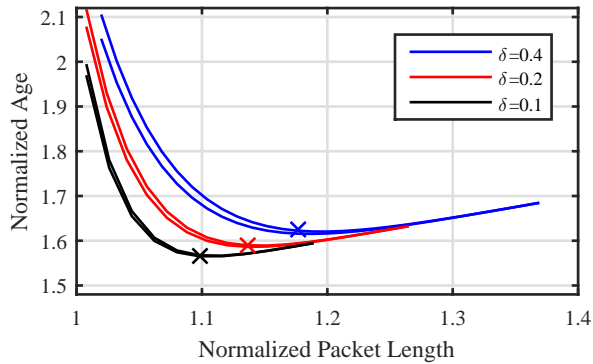}
\caption{The average age $\FRage{n}$ and the upper bound $\FRUB{n}$ for update packets with $k=50$ information symbols as a function of the number of transmitted symbols $n$. The age and packet length are normalized by $k/(1-\delta)$. For each curve, $\times$ marks $\FRUB{\nhat^*_k}$.}\label{fig:FRUB50}\negfigspace
\end{figure}

Figure~\ref{fig:FRUB50} is similar to Figure~\ref{fig:FR1000} except there are only $k=50$ information bits and the figure includes the upper bound $\FRUB{n}$ in \eqnref{AgeQ-UB}. Furthermore, the figure is plotted as the normalized age $\age/[k/(1-\delta)]$ vs.~the normalized packet length $n/[k/(1-\delta)]$. This normalization and the small value of $k$ are chosen to accentuate the gap between $\FRage{n}$ and $\FRUB{n}$.  For typical values of $k$ such as $k=1000$, the gap between the age and the upper bound cannot be visually resolved.

\begin{figure}[t]
\centering
\includegraphics{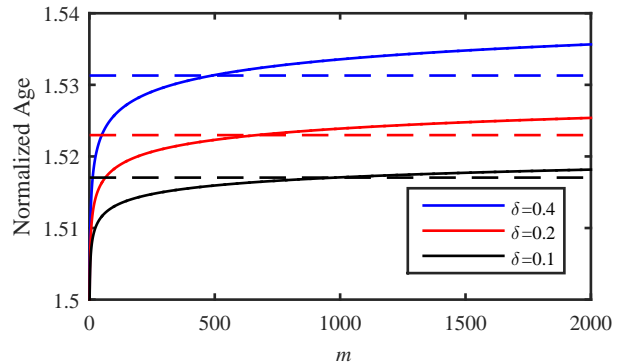}
\caption{The average age $\ageIIR^{(m)}$ (solid line) and the upper bound $\FRUB{\nhat^*_k}$ (dashed line) for update packets with $k=1000$ information symbols as a function of the number of monitors $m$. Note that  ages are normalized by $k/(1-\delta)$.}\label{fig:IIRm}\negfigspace
\end{figure}

Figure~\ref{fig:IIRm} compares systems with $m>1$ monitors.  
For each value of $\delta$, we compare the IIR age $\ageIIR^{(m)}$ and the FR age $\FRage{\nhat^*_k}$ using $\nhat^*_k$ symbols matched to the erasure rate $\delta$.  The IIR system completes the transmission of an update only after all $m$ monitors have decoded. Consequently, the IIR average age grows monotonically with $m$. Because the FR system ignores whether a monitor has actually decoded an update, the FR age is insensitive to the number of users. To highlight small differences, all ages are normalized by $k/(1-\delta)$.  We see that for all values of the erasure probability $\delta$, the FR system outperforms the IIR when the number of monitors $m$ becomes sufficiently large. We also see that normalized system performance is very similar across a range of erasure rates.

\section{Discussion}\label{sec:conclusion}
We have shown that the FR system, which requires no feedback, can essentially match the performance of the IIR system that does require update delivery feedback from the monitor. However, the FR system does require the redundancy to be carefully optimized in response to the channel erasure rate. In practical systems,  the erasure rate  will vary with time and cannot be assumed to be known. Hence, the FR system will also require some form of feedback to establish the appropriate redundancy level.  In practice, systems issues, such as whether receiver feedback can be supported, will determine  which approach is better in a particular setting.

In addition, other coded redundancy mechanisms merit examination. 
For example, the finite incremental redundancy  (FIR) strategy \cite{heindlmaier2014isn}, just like FR, uses a fixed rate code, but the source is provided instantaneous feedback if the update has been decoded before all its $n$ symbols have been transmitted, at which point it starts transmitting a new update, as in IIR.  On the other hand, FIR shares the advantage of FR that updates that were unlucky in transmission can be terminated without waiting for successful decoding. 


It is natural to compare timely update delivery with HARQ-aided content download.
Hybrid ARQ (HARQ) is a special transmission scheme that combines the conventional ARQ with error correction (see e.g.~\cite{HARQ:SoljaninLS03}). Incremental redundancy HARQ (IR-HARQ) schemes adapt their error correcting code redundancy to varying channel conditions, and thus achieve better throughput efficiency than ordinary ARQ. 
In content download, all content needs to be delivered, and thus these systems have to have a rateless transmission at some level (e.g., conventional ARQ or Fountain codes at the packet level) which will continue until each packet is successfully delivered. For example, in eMBMS, an FR strategy on the physical layer would have a Fountain code at the packet level \cite{shokrollahi2011raptor}. Furthermore, content download systems strive to minimize the download time, which, as we have seen in Sec~\ref{sec:IIR}, is not equivalent to minimizing AoI.  The behavior  of both systems in a multi-user scenario is similar because of an
underlying order statistics phenomenon. Roughly speaking, when there are many users, it is very likely that it will take a long time for some to decode, and putting limits on that time as the FR strategy does, will have an advantage. Content download systems will then have to supplement such systems with an outer rateless code. It would be interesting to compare update delivery with content streaming where all packets have to be delivered  in a timely manner.


\section*{Acknowledgements}

This research was supported in part by grant No.  200021\_166106/1 of the Swiss National Science Foundation and in part by the NSF Award CIF-1422988.

\bibliographystyle{IEEEtran}
\bibliography{IEEEabrv,ehcache}

\appendix
\begin{IEEEproof} (Lemma~\ref{lem:rn}) With the shorthand definitions 
\begin{align}\eqnlabel{q-qhat-defn}
q_n&\equiv\frac{\Fk{n}}{(1-\delta)^k}, 
& \qhat_n&\equiv\frac{\Fk[k+1]{n+1}}{(1-\delta)^{k+1}},
\end{align}
 we observe that \eqnref{mutil} permits us to write
\begin{align}\eqnlabel{mutil-ratio}
\mutil_n&=k\qhat_n/q_n. 
\end{align}
It follows from \eqnref{X-CDF} and \eqnref{q-qhat-defn} that $q_k=\qhat_k=1$ and that
\begin{align}
q_{k+1}&=1+k\delta, & \qhat_{k+1}&= 1+(k+1)\delta.
\end{align}
These facts imply $\mutil_k=k$ and 
\begin{align}
\mutil_{k+1}&= \frac{k\qhat_{k+1}}{q_{k+1}}=k+\frac{k\delta}{1+k\delta}.
\end{align}
Thus $\mutil_k\le \mutil_{k+1}$ and $\mutil_{k+1}\le k+1$. We now prove by induction that the sequence $\mutil_n$ is nondecreasing and satisfies $\mutil_n\le n$.  Suppose $\mutil_k\le \mutil_{k+1}\le \cdots \le \mutil_{n-1}$ and that $\mutil_i\le i$ for $i<n$.
Defining
$\gamma_n\equiv\binom{n-1}{k-1}\delta^{n-k}$,
it follows from \eqnref{X-CDF} that
\begin{align}
q_n&=q_{n-1} +\gamma_n,\\
\qhat_n&=\qhat_{n-1} +\binom{n}{k}\delta^{n-k}
=\qhat_{n-1}+\frac{n}{k}\gamma_n.
\end{align}
This implies
\begin{align}\eqnlabel{rn2}
\mutil_n=\frac{k\qhat_n}{q_n}
&=\frac{k\qhat_{n-1}+n\gamma_{n}}{q_{n-1}+\gamma_n}.
\end{align}
By our induction hypothesis, $\mutil_{n-1}\le n$, or, equivalently, $k\qhat_{n-1}\le nq_{n-1}$. Applying this upper bound to the numerator in 
\eqnref{rn2} yields $\mutil_n\le n$.
We now observe that \eqnref{rn2} and 
$n\ge \mutil_{n-1}=k\qhat_{n-1}/q_{n-1}$ also imply
\begin{align}\eqnlabel{rn-inc}
\mutil_n
&\ge\frac{k\qhat_{n-1}+(k\qhat_{n-1}/q_{n-1})\gamma_{n}}{q_{n-1}+\gamma_n}=\frac{k\qhat_{n-1}}{q_{n-1}}=\mutil_{n-1}.
\end{align}
Finally, we observe from \eqnref{XkCDF} that
\begin{align}
\limty{n} q_n&=\frac{1}{(1-\delta)^k},
&\limty{n}\qhat_n&=\frac{1}{(1-\delta)^{k+1}}.
\end{align}
This implies $\limty{n} \mutil_n =k/(1-\delta)$. Since $\mutil_n$ is nondecreasing, $\mutil_n\le k/(1-\delta)$ for all $n\ge k$. 
\end{IEEEproof}

\begin{IEEEproof} (Lemma~\ref{lem:Xktail}) Random variable $\Xk$ has moment generating function $\phi_{\Xk}(s)=[(1-\delta)e^{s}/(1-\delta e^s)]^k$. 
By the Chernoff bound, $\ln\prob{\Xk\ge \nhat^*_k}\le \min_{s\ge0} \Pk(s)$ where
\begin{align}
\Pk(s)&=\ln[e^{-s\nhat^*}\phi_{\Xk}(s)]\\
&=k\bracket{\ln(1-\delta)-s\parfrac{\delta+\wk}{1-\delta}-\ln(1-\delta e^s)}.
\end{align}
It is straightforward to show that $P_k(s)$ is minimized at 
\begin{align}\eqnlabel{s*}
s^*=\ln[(1+\wk/\delta)/(1+\wk)].
\end{align}
Using the shorthand notation
$\LX(x)=\ln(1+x)$,
it follows from \eqnref{s*} that
\begin{align}
\Pk(s^*)=\frac{-k%
\bracket{(\delta+\wk)\LX(\wk/\delta)-(1+\wk)\LX(\wk)}}{1-\delta}.
\end{align}
Defining 
\begin{align}
y_1(k) &\equiv k[\delta\LX(\wk/\delta)-\LX(\wk)],\\
y_2(k)&\equiv k\wk\bracket{\LX(\wk/\delta)-\LX(\wk)},
\end{align}
we observe that 
\begin{align}\eqnlabel{Pk2}
\Pk(s^*)=-\frac{y_1(k)+y_2(k)}{1-\delta}.
\end{align}
With the definition
\begin{align}
\lk\equiv\ln\parfrac{2k}{\pi\delta}=\ln k+\ln\parfrac{2}{\pi\delta},
\end{align}
we observe from \eqnref{ekdefn} that $\wk^2=\delta\lk/k$.
This implies $y_1(k) =\lk R_1(k)$ and $y_2(k)=\lk R_2(k)$ where
\begin{align}
R_1(k)&=\frac{\delta\bracket{\delta\LX(\wk/\delta)-\LX(\wk)}}{\wk^2},\\
R_2(k)&=\frac{\delta\bracket{\LX(\wk/\delta)-\LX(\wk)}}{\wk}.
\end{align}
Since $\wk\to0$ as $k\to\infty$,  l'H\^opital's rule yields
\begin{align}\eqnlabel{R1lim}
\limty{k}R_1(k)&=\lim_{z\to0}\frac{\delta\bracket{\delta\LX(z/\delta)-\LX(z)}}{z^2}
= -\frac{1-\delta}{2},\\
\limty{k}R_2(k)&=\limty{z}\frac{\delta\bracket{\LX(z/\delta)-\LX(z)}}{z} =
1-\delta.\eqnlabel{R2lim}
\end{align}
It follows from \eqnref{Pk2} that
\begin{align}
\Pk(s^*)=-\frac{\lk}{1-\delta}[R_1(k)+R_2(k)].
\end{align}
Moreover, \eqnref{R1lim} and \eqnref{R2lim} imply that for any $\eta_0>0$, there exists $K_0$ such that 
\begin{align}
\Pk(s^*)=-\frac{\lk}{1-\delta}\paren{\frac{1-\delta}{2}-\eta_0},\quad
k\ge K_0.
\end{align}
The claim then follows.
\end{IEEEproof}

%

%



\end{document}